\documentclass[aps,preprint,amsmath,amssymb]{revtex4}
\usepackage{graphicx}
\usepackage{epstopdf}
\begin{document}

\title{Probe of the anomalous quartic couplings with beam polarization at the CLIC}

\author{A. Senol}
\email[]{senol_a@ibu.edu.tr} \affiliation{Department of Physics,
Abant Izzet Baysal University, 14280, Bolu, Turkey}

\author{M. K\"{o}ksal}
\email[]{mkoksal@cumhuriyet.edu.tr} \affiliation{Department of
Optical Engineering, Cumhuriyet University, 58140, Sivas, Turkey}

\author{S. C. \.{I}nan}
\email[]{sceminan@cumhuriyet.edu.tr} \affiliation{Department of Physics, Cumhuriyet University, 58140, Sivas, Turkey}

\begin{abstract}
We have investigated the anomalous quartic couplings defined by the dimension-8 operators in semi-leptonic decay channel of the  $e^{+}e^{-}\rightarrow \nu_{e} W^{-} W^{+} \bar{\nu}_{e}$ process for unpolarized and polarized electron (positron) beam at the Compact Linear Collider. We give the $95\%$ confidence level  bounds on the anomalous $\frac{f_{S0}}{\Lambda^{4}}$, $\frac{f_{S1}}{\Lambda^{4}}$ and $\frac{f_{T0}}{\Lambda^{4}}$ couplings for various values of the integrated luminosities and center-of-mass energies. The best sensitivities obtained on anomalous $\frac{f_{S0}}{\Lambda^{4}}$, $\frac{f_{S1}}{\Lambda^{4}}$ and $\frac{f_{T0}}{\Lambda^{4}}$ couplings through the process $e^{+}e^{-}\rightarrow \nu_{e} W^{-} W^{+} \bar{\nu}_{e}$ with beam polarization at $\sqrt{s}=3$ TeV and an integrated luminosity of $L_{int}=2000$ fb$^{-1}$ are $[-4.05;\, 3.67]\times 10^{-12}$ GeV$^{-4}$, $[-3.08; \, 2.12]\times 10^{-12}$ GeV$^{-4}$, $[-1.98; \, 0.64]\times 10^{-13}$ GeV$^{-4}$, which show improvement over the current bounds.
\end{abstract}

\maketitle

\section{Introduction}

The Standard Model (SM) has been  proven to be highly successful through many significant experimental tests, in particular the discovery of a new particle consistent with the SM Higgs boson with a mass between $125-126$ GeV detected by the ATLAS and the CMS experiments at the LHC \cite{higgs1,higgs2}. On the other hand, since many important questions, such as the origin of mass, the large hierarchy between electroweak and the Planck scale, the strong CP problem, and the matter/antimatter asymmetry remain unanswered in the SM, we need to study physics beyond the SM. One of the ways of probing new physics beyond the SM is to investigate the anomalous gauge boson interactions. Gauge boson self-interactions in the SM are exactly described by the $SU_{L}(2)\times U_{Y}(1)$ gauge symmetry. The precision measurements of gauge boson self-interactions can further verify the SM. Furthermore, the existence of anomalous gauge boson couplings
may be a sign of new physics beyond the SM. The effective Lagrangian approach is one of the common ways for searching new physics beyond the SM in a model independent way. In particular, the anomalous quartic gauge boson couplings can be examined with the aid of the effective Lagrangian approach. Such an approach is parameterized by high-dimensional operators which induce anomalous quartic gauge couplings that modify the interactions between the electroweak gauge bosons.

The LHC is expected to reply some of the fundamental open questions in particle physics.
Nevertheless, the analysis of the LHC data is quite difficult due to remnants of the usual proton-proton deep inelastic processes.
Whereas, collisions between electrons and positrons are much simpler to investigate than proton-proton collisions.
A linear electron-positron collider with high luminosity and
energy is the best option to complement and to expand the LHC physics program. The
CLIC is one of the most popular linear colliders, purposed to follow out electron-positron collisions at
energies from 0.35 TeV to 3 TeV \cite{cillic}. To have its high luminosity and energy is quite important
with regards to new physics research beyond the SM. Since the anomalous quartic couplings defining through effective Lagrangians
have dimension-8, they have very strong energy dependence. Therefore, the anomalous cross section including these vertices has higher energy dependence than the SM cross section. Hence, CLIC will have a great potential to examine the anomalous quartic
gauge boson couplings.

High-dimensional effective operators describing the anomalous quartic gauge boson couplings are expressed by either linear or
nonlinear effective Lagrangians. Nonlinear effective Lagrangians are considered if there is no
Higgs boson in the low energy spectrum. However, linear effective Lagrangians are obtained by using a linear representation of gauge symmetry that is broken by the conventional SM Higgs mechanism. It becomes important to study the anomalous quartic gauge boson couplings based on linear effective Lagrangians due to the discovery of a Higgs boson in the LHC. For these reasons, we only deal with dimension-8 operators in our work.

In this paper, we will analyze the anomalous quartic couplings via $e^{+}e^{-}\rightarrow \nu_{e} W^{-} W^{+} \bar{\nu}_{e}$ process with semi-leptonic decay including polarized electron (positron) beam effects at the CLIC for the center-of-mass energies of 1.4 and 3 TeV.
\section{Dimension-eight operators for quartic gauge couplings}

There are three classes of  operators that describe the anomalous quartic couplings. The first class of operators can be parameterized in terms of only the covariant
derivative of the field $D_{\mu}\Phi$. This class includes two independent operators \cite{baak}:
\begin{eqnarray}
L_{S0}&=&\frac{f_{S0}}{\Lambda^{4}}[(D_{\mu}\Phi)^{\dag}D_{\nu}\Phi]\times[(D^{\mu}\Phi)^{\dag}D^{\nu}\Phi],\\
L_{S1}&=&\frac{f_{S1}}{\Lambda^{4}}[(D_{\mu}\Phi)^{\dag}D_{\mu}\Phi]\times[(D^{\nu}\Phi)^{\dag}D^{\nu}\Phi],
\end{eqnarray}
The second class of operators are related to $D_{\mu}\Phi$ and the field strength. These seven operators are given as follows \cite{baak}
\begin{eqnarray}
L_{M0}&=&\frac{f_{M0}}{\Lambda^{4}}Tr[W_{\mu\nu}W^{\mu\nu}]\times[(D_{\beta}\Phi)^{\dag}D^{\beta}\Phi],\\
L_{M1}&=&\frac{f_{M1}}{\Lambda^{4}}Tr[W_{\mu\nu}W^{\nu\beta}]\times[(D_{\beta}\Phi)^{\dag}D^{\mu}\Phi],\\
L_{M2}&=&\frac{f_{M2}}{\Lambda^{4}}[B_{\mu\nu}B^{\mu\nu}]\times[(D_{\beta}\Phi)^{\dag}D^{\beta}\Phi],\\
L_{M3}&=&\frac{f_{M3}}{\Lambda^{4}}[B_{\mu\nu}B^{\nu\beta}]\times[(D_{\beta}\Phi)^{\dag}D^{\mu}\Phi],\\
L_{M4}&=&\frac{f_{M4}}{\Lambda^{4}}[(D_{\mu}\Phi)^{\dag}W_{\beta\nu}D^{\mu}\Phi)]\times B^{\beta\nu},\\
L_{M5}&=&\frac{f_{M5}}{\Lambda^{4}}[(D_{\mu}\Phi)^{\dag}W_{\beta\nu}D^{\nu}\Phi)]\times B^{\beta\mu},\\
L_{M6}&=&\frac{f_{M6}}{\Lambda^{4}}[(D_{\mu}\Phi)^{\dag}W_{\beta\nu}W^{\beta\nu}D^{\mu}\Phi)],\\
L_{M7}&=&\frac{f_{M7}}{\Lambda^{4}}[(D_{\mu}\Phi)^{\dag}W_{\beta\nu}W^{\beta\mu}D^{\nu}\Phi)],
\end{eqnarray}
The remaining operators contain, solely, the field strength tensors. These operators can be expressed as \cite{baak}
\begin{eqnarray}
L_{T0}&=&\frac{f_{T0}}{\Lambda^{4}}Tr[W_{\mu\nu}W^{\mu\nu}]\times Tr[W_{\alpha\beta}W^{\alpha\beta}],\\
L_{T1}&=&\frac{f_{T1}}{\Lambda^{4}}Tr[W_{\alpha\nu}W^{\mu\beta}]\times Tr[W_{\mu\beta}W^{\alpha\nu}],\\
L_{T2}&=&\frac{f_{T2}}{\Lambda^{4}}Tr[W_{\alpha\mu}W^{\mu\beta}]\times Tr[W_{\beta\nu}W^{\nu\alpha}],\\
L_{T5}&=&\frac{f_{T5}}{\Lambda^{4}}Tr[W_{\mu\nu}W^{\mu\nu}]\times B_{\alpha\beta}B^{\alpha\beta},\\
L_{T6}&=&\frac{f_{T6}}{\Lambda^{4}}Tr[W_{\alpha\nu}W^{\mu\beta}]\times B_{\mu\beta}B^{\alpha\nu},\\
L_{T7}&=&\frac{f_{T7}}{\Lambda^{4}}Tr[W_{\alpha\mu}W^{\mu\beta}]\times [B_{\beta\nu}B^{\nu\alpha}],\\
L_{T8}&=&\frac{f_{T8}}{\Lambda^{4}}B_{\mu\nu}B^{\mu\nu}B_{\alpha\beta}B^{\alpha\beta},\\
L_{T9}&=&\frac{f_{T9}}{\Lambda^{4}}B_{\alpha\mu}B^{\mu\beta}B_{\beta\nu}B^{\nu\alpha}.
\end{eqnarray}

\noindent The complete list of quartic vertices modified by these operators is given in Table \ref{tab1}.

\begin{figure}
\includegraphics{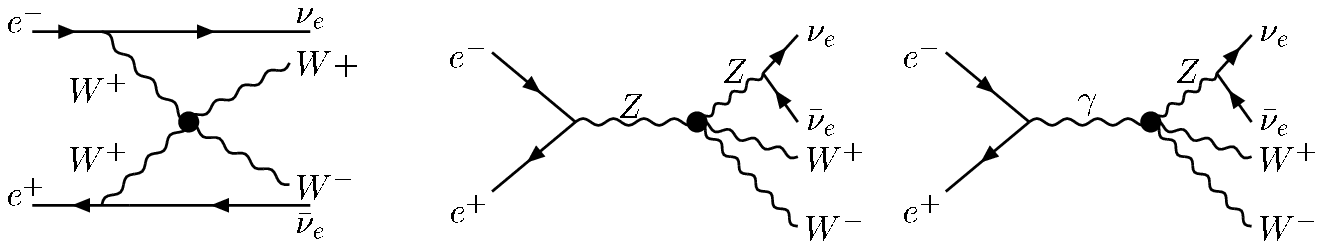}
\caption{The Feynman diagrams for the $e^{+}e^{-}\rightarrow \nu_{e} W^{-} W^{+} \bar{\nu}_{e}$ process in the presence of anomalous quartic $WWWW$, $WWZ\gamma$ and $WWZZ$ couplings.
\label{fig0}}
\end{figure}

There are total 59 Feynman diagrams for process $e^+e^-\to W^+ W-\nu_e\bar\nu_e$ including  anomalous quartic $WWWW$, $WWZ\gamma$ and $WWZZ$ couplings.  The three diagrams in presence of anomalous quartic $WWWW$, $WWZ\gamma$ and $WWZZ$ couplings are shown in Fig \ref{fig0}. We can see from Table I that while $L_{S0}$ and $L_{S1}$ operators modify the anomalous $WWWW$ and $WWZZ$ couplings, dimension-$8$ effective $L_{T0}$ operator causes the anomalous $WWWW$, $WWZ\gamma$ and $WWZZ$ vertices.

There have been many studies for the anomalous gauge self-interactions at linear and
hadron colliders. On the other hand, the anomalous quartic couplings arising from dimension-8 operators at the LHC and the future hadron colliders have been investigated in Refs. \cite{11,12,13,14,15,16,17,18,19,20,21,22,23,24,25,26,27,28,29,30,31,32,33,34,35,36,37,38,39,41,344,lhc1,lhc2,lhc3, fab}.  In our paper, we choose $L_{S0}$, $L_{S1}$ and $L_{T0}$ operators to investigate anomalous quartic couplings. In the literature, these couplings have been examined three different $\frac{f_{S0}}{\Lambda^{4}}$, $\frac{f_{S1}}{\Lambda^{4}}$ and $\frac{f_{T0}}{\Lambda^{4}}$ couplings.
Ref.\cite{344} has been experimentally obtained the bounds on the $\frac{f_{S0}}{\Lambda^{4}}$ and $\frac{f_{S1}}{\Lambda^{4}}$ using proton-proton collision data corresponding to an integrated luminosity of $20.3$ fb$^{-1}$ at a center-of-mass energy of 8 TeV collected by the ATLAS detector at the LHC. In Ref.\cite{lhc1}, the sensitivity bounds are obtained only on $\frac{f_{S0}}{\Lambda^{4}}$ and  $\frac{f_{S1}}{\Lambda^{4}}$ at $99\%$ confidence level (C.L.) for $\sqrt{s}=14$ TeV via $pp \to jj e^{\pm}\mu^{\pm} \nu \nu$ process. Similarly the sensitivity bounds on $\frac{f_{T0}}{\Lambda^{4}}$ with $95\%$ C.L. via triboson production at the proton-proton colliders for $\sqrt{s}=14$ TeV  are obtained in Ref.\cite{lhc2}. In Ref.\cite{lhc3}, the $\frac{f_{S0}}{\Lambda^{4}}$, $\frac{f_{S1}}{\Lambda^{4}}$ and $\frac{f_{T0}}{\Lambda^{4}}$ couplings have been studied  via $WWW$ final state with full leptonic decay and semi-leptonic decay for $\sqrt{s}=14$ TeV at the LHC and $\sqrt{s}=100$ TeV for future hadron collider. The best available constraints on $\frac{f_{S0}}{\Lambda^{4}}$, $\frac{f_{S1}}{\Lambda^{4}}$ and $\frac{f_{T0}}{\Lambda^{4}}$ parameters defining these anomalous quartic couplings obtained from one parameter analysis in Refs.\cite{lhc1,lhc2,lhc3,344} are summarized in Table \ref{tab2}. In reference \cite{fab} the authors examine the non-linear parametrization through the processes $pp\rightarrow l^\pm \nu_l l^\pm \nu_l jj$ and $pp\rightarrow l^\pm \nu_l l^\mp \nu_l jj$ at the LHC.

\section{Numerical Analysis}

The important features of a linear collider are its clean experimental
environment, high energy and polarized beams. A polarized electron beam
would provide suitable platform searching of the SM and for
diagnosing new physics. Observation of even the tiniest signal which conflicts with the SM expectations would be a convincing evidence for physics beyond the SM. Proper selections of the electron and positron beam polarizations
may therefore be used to enhance the new physics signal and also to considerably suppress backgrounds.

We use MadGraph5 \cite{mdg} to generate the signal and background events with the effective Lagrangian implemented through FEYNRULES \cite{feyn}.  In order to probe the sensitivity of anomalous quartic couplings ($L_{S0}$, $L_{S1}$ and $L_{T0}$), we analyzed the process $e^+e^-\to W^+(\to l^+\nu) W^-(\to jets) \nu_e\bar{\nu_e}$. In our analysis the main background processes yield identical final states to the signal process.  Here we assume that one of the $W$'s in the final state decays into leptonic channel and the other one decays into hadronic channel. Therefore, the final state signal and background topology of the process consists of an energetic lepton ($l$), neutrinos (missing ${E}_T$), and two hadronic jets ($2j$). We apply following set of cuts in order to suppress the backgrounds and enhance the signal for the anomalous quartic interactions in $e^+e^-\to W^+ W \nu_e\bar\nu_e$ process.

(1) $P_{Tj}>10$ GeV, $|\eta_j|<5$

(2) $E_{miss}>10$ GeV (missing $E_T$ (sum of neutrino's momenta)),

(3) $|M_{jj}-M_W|<15$ GeV,

(4) $\Delta R>0.4$,

\noindent where $j=u,d,s,c$ and $\Delta R$ is the angular separation between any two objects. $\Delta R=\sqrt{\Delta \varphi^2+ \Delta \eta^2 }$ with $\varphi$ representing the azimuthal angle with respect to the beam directions.

\begin{figure}
\includegraphics{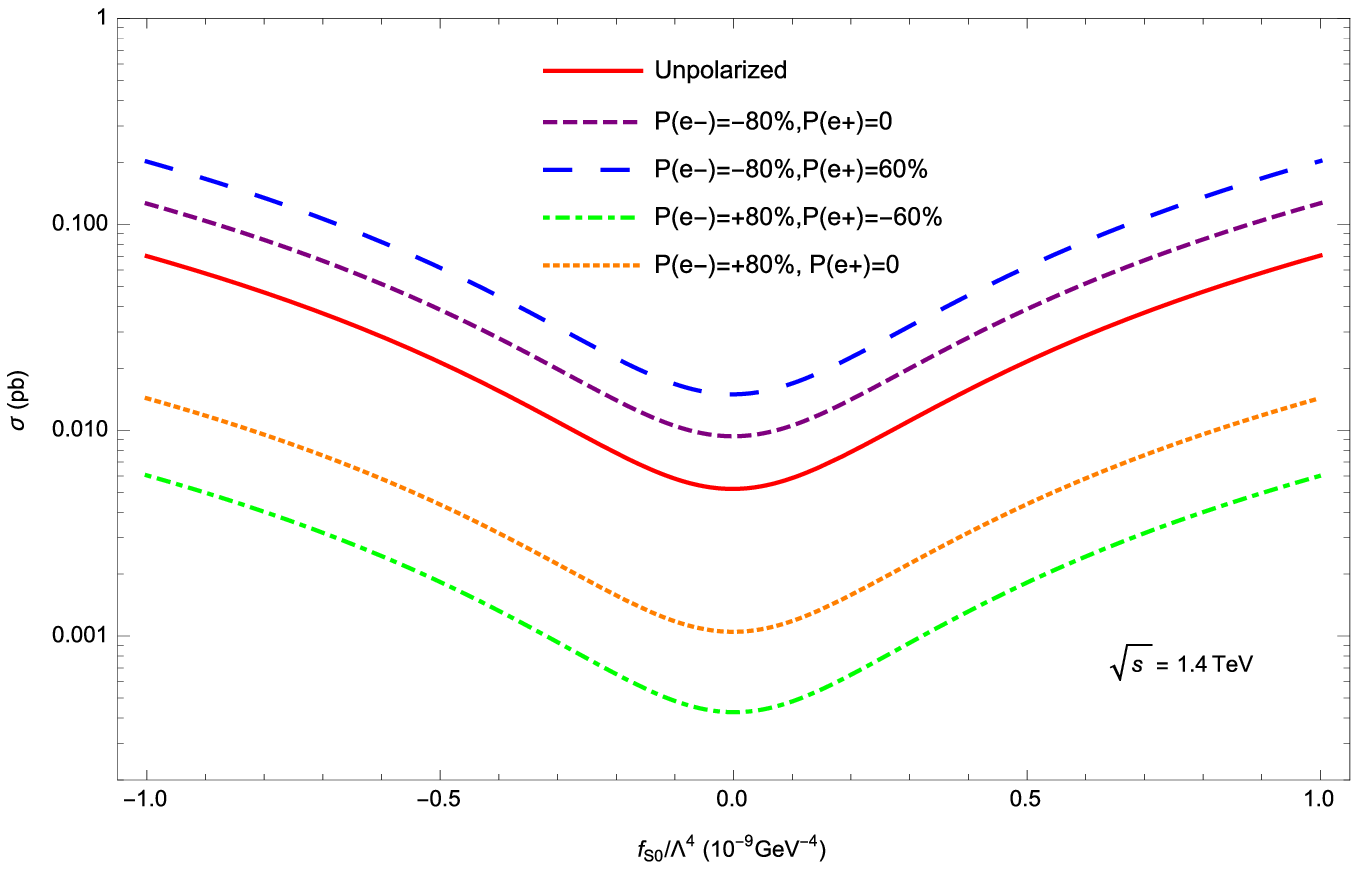}
\caption{The total cross section for $e^{+}e^{-}\rightarrow \nu_{e} W^{-} W^{+} \bar{\nu}_{e}$ processes as a function of $\frac{f_{S0}}{\Lambda^{4}}$ at the $\sqrt{s}=1.4$ TeV for different polarisation of the positron and electron beams.
\label{fig1}}
\end{figure}

\begin{figure}
\includegraphics{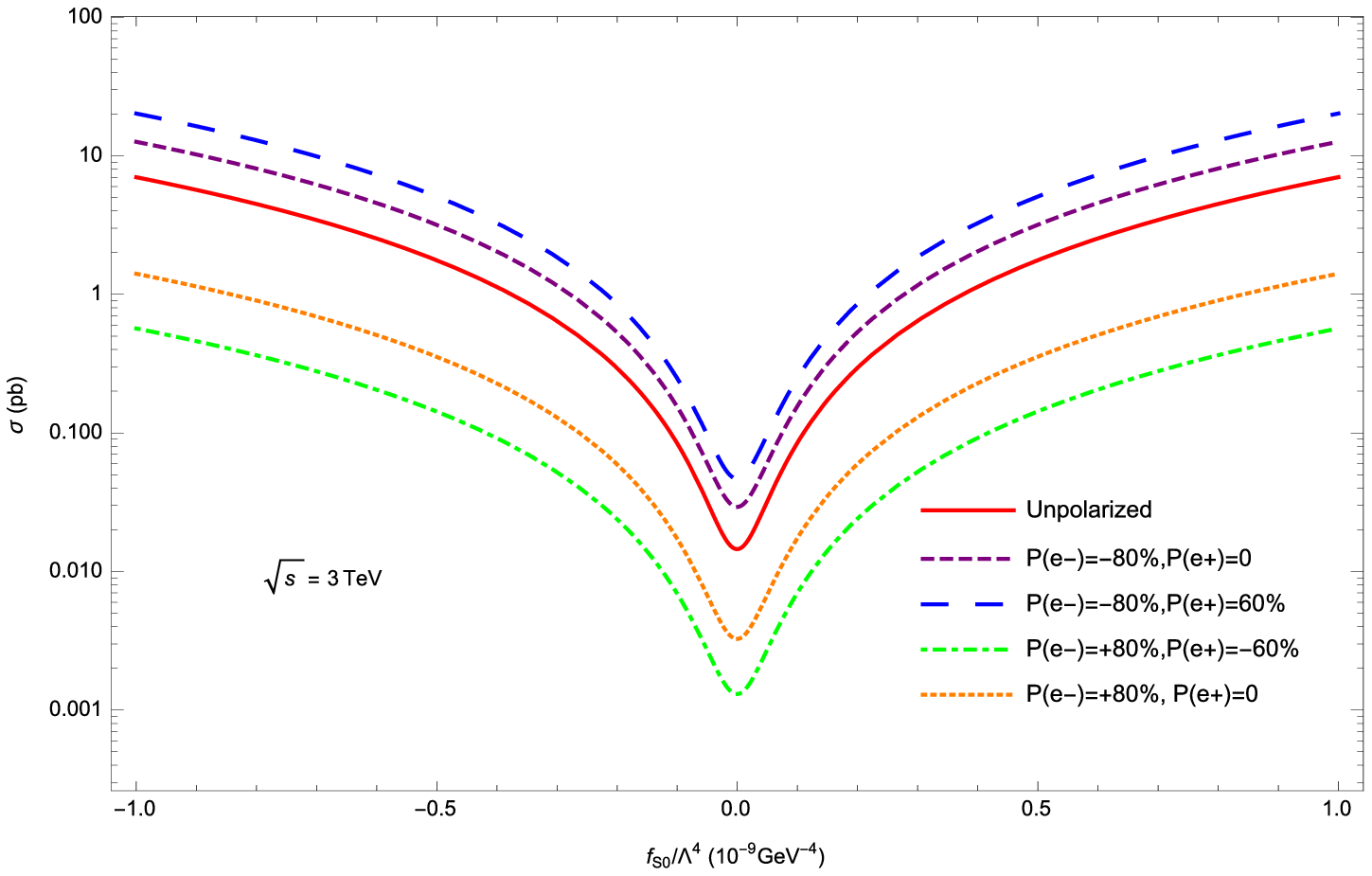}
\caption{The total cross section for $e^{+}e^{-}\rightarrow \nu_{e} W^{-} W^{+} \bar{\nu}_{e}$ processes as a function of $\frac{f_{S0}}{\Lambda^{4}}$ at the $\sqrt{s}=3$ TeV for different polarisation of the positron and electron beams.
\label{fig2}}
\end{figure}

\begin{figure}
\includegraphics{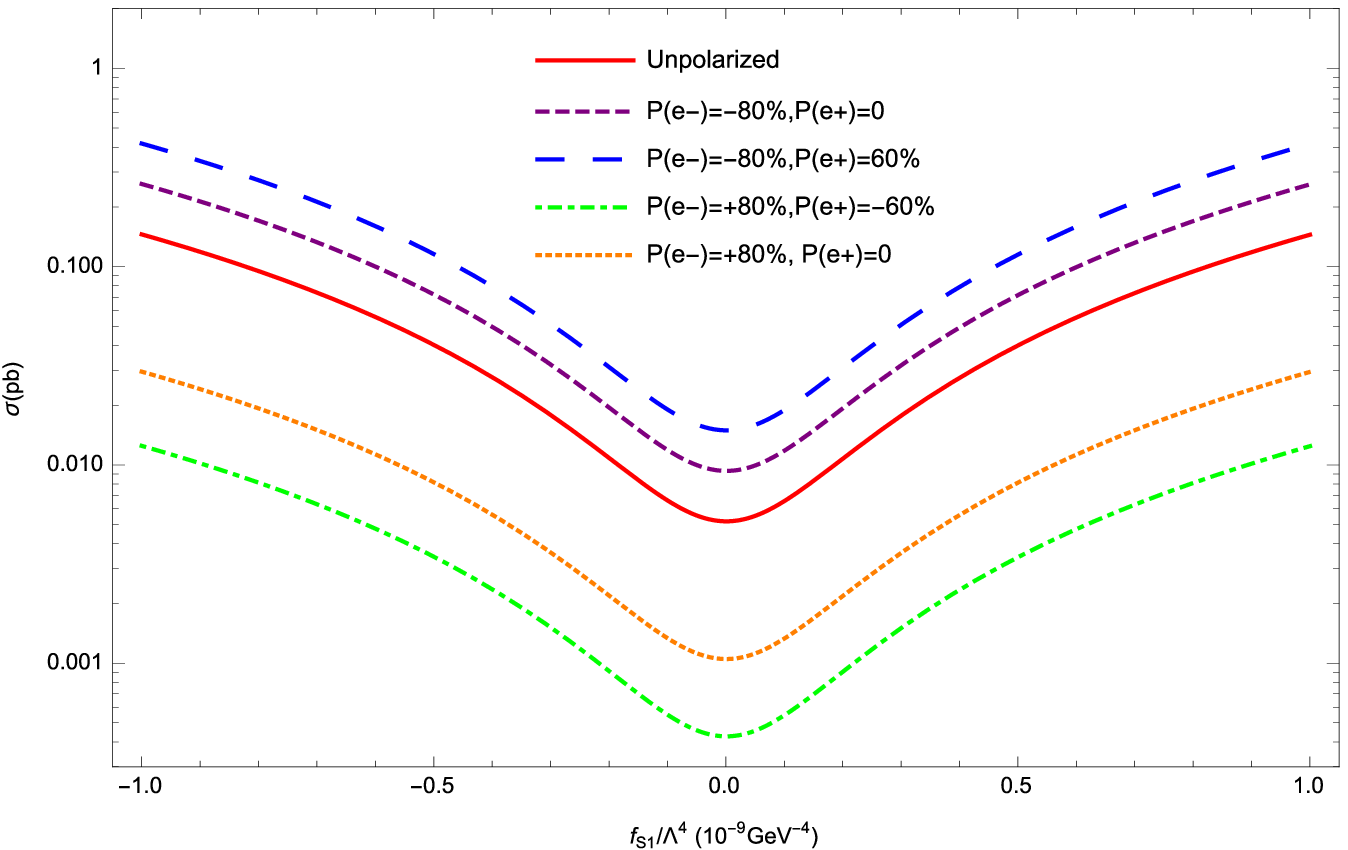}
\caption{The total cross section for $e^{+}e^{-}\rightarrow \nu_{e} W^{-} W^{+} \bar{\nu}_{e}$ processes as a function of $\frac{f_{S1}}{\Lambda^{4}}$ at the $\sqrt{s}=1.4$ TeV for different polarisation of the positron and electron beams.
\label{fig3}}
\end{figure}

\begin{figure}
\includegraphics{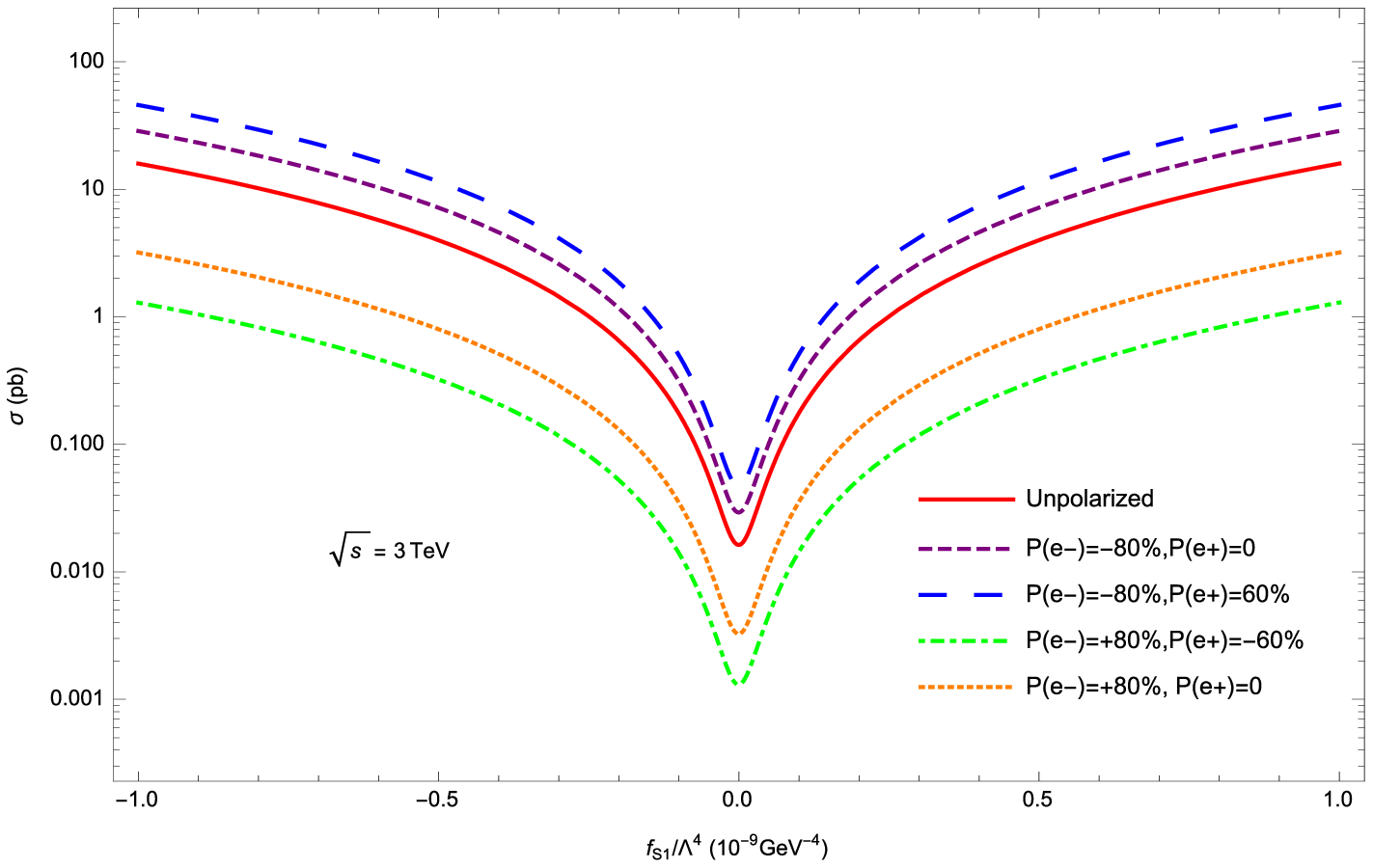}
\caption{The total cross section for $e^{+}e^{-}\rightarrow \nu_{e} W^{-} W^{+} \bar{\nu}_{e}$ processes as a function of $\frac{f_{S1}}{\Lambda^{4}}$ at the $\sqrt{s}=3$ TeV for different polarisation of the positron and electron beams.
\label{fig4}}
\end{figure}

\begin{figure}
\includegraphics{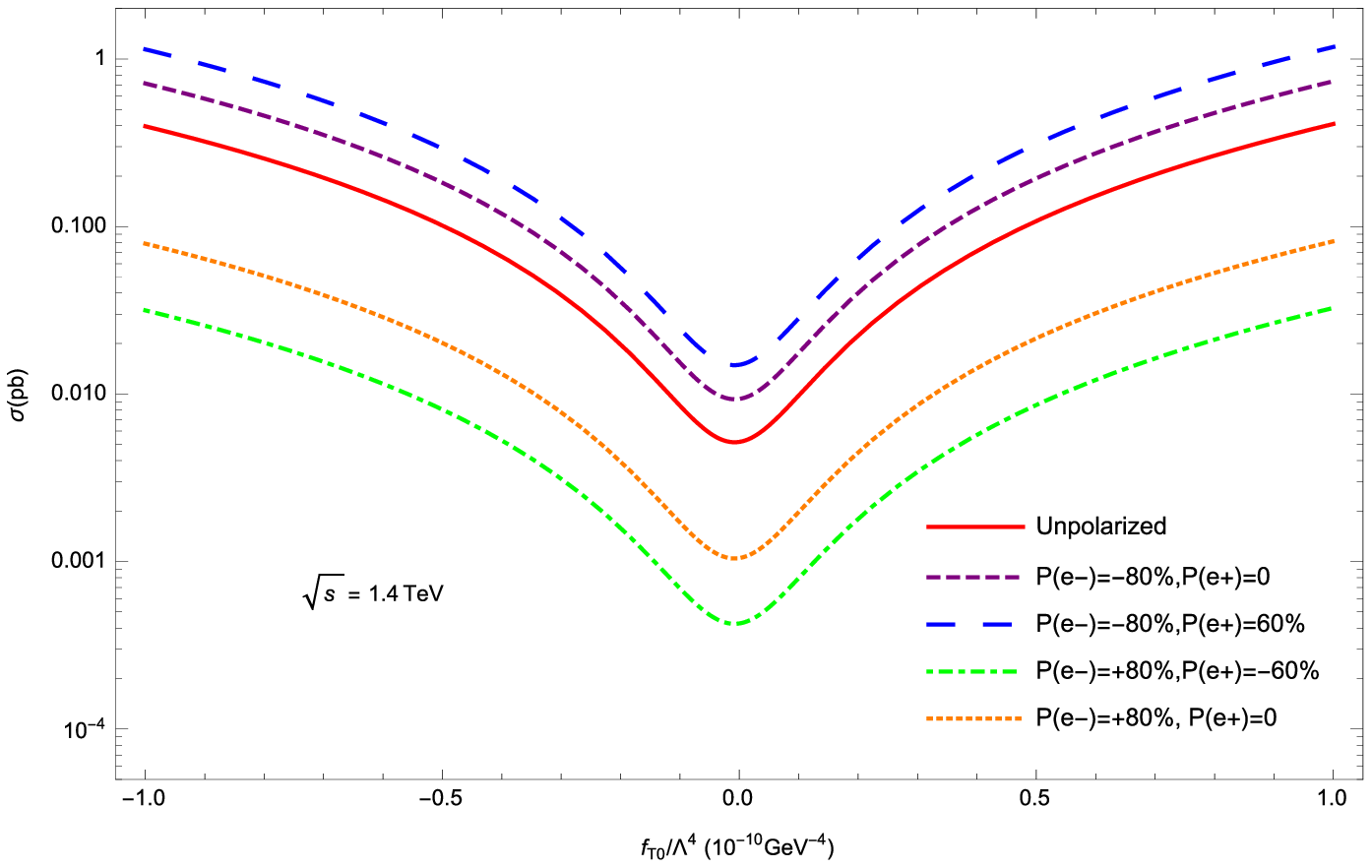}
\caption{The total cross section for $e^{+}e^{-}\rightarrow \nu_{e} W^{-} W^{+} \bar{\nu}_{e}$ processes as a function of $\frac{f_{T0}}{\Lambda^{4}}$ at the $\sqrt{s}=1.4$ TeV for different polarisation of the positron and electron beams.
\label{fig5}}
\end{figure}

\begin{figure}
\includegraphics{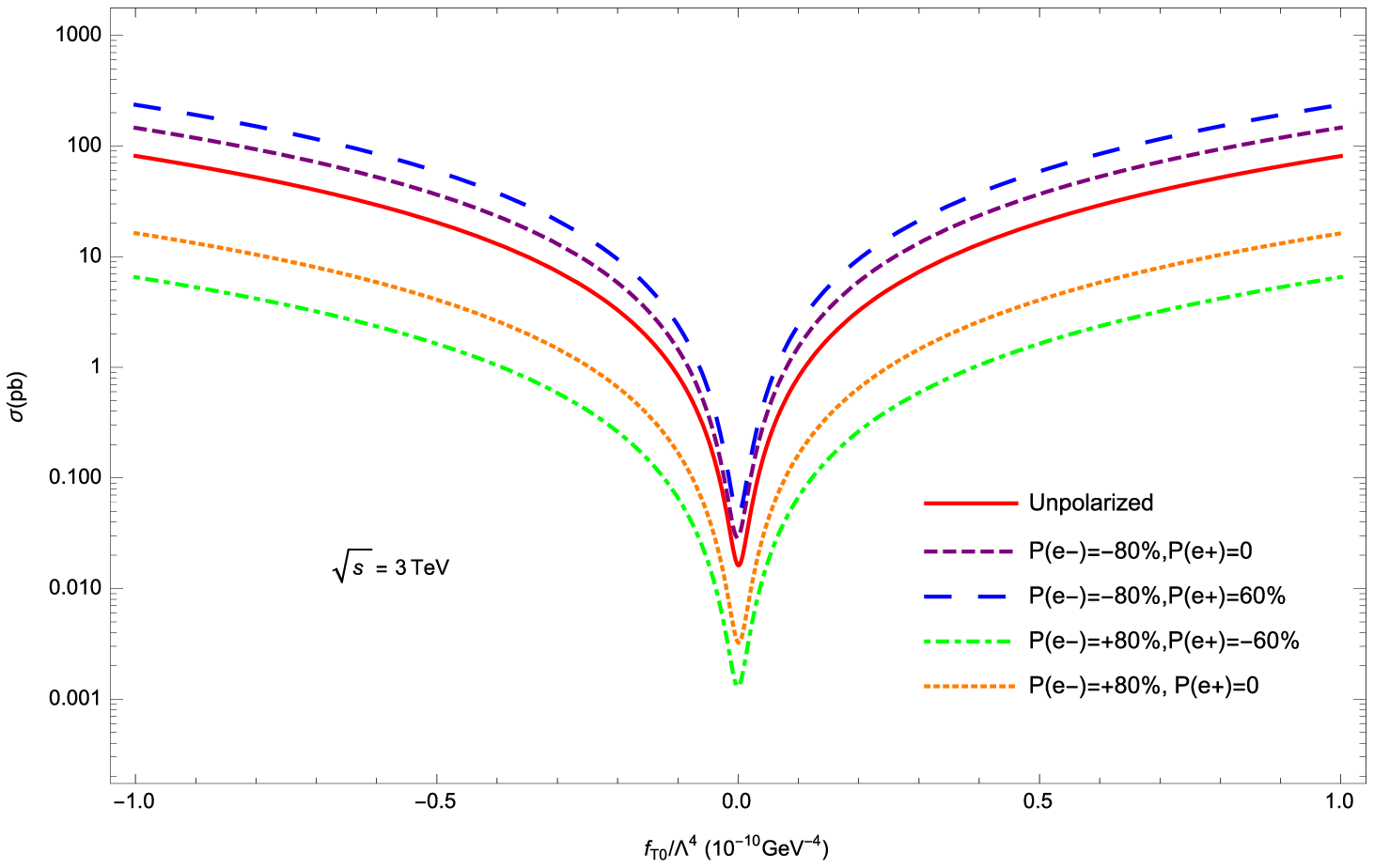}
\caption{The total cross section for $e^{+}e^{-}\rightarrow \nu_{e} W^{-} W^{+} \bar{\nu}_{e}$ processes as a function of $\frac{f_{T0}}{\Lambda^{4}}$ at the $\sqrt{s}=3$ TeV for different polarisation of the positron and electron beams.
\label{fig6}}
\end{figure}

If the electroweak nature of the interactions of the processes are taken into account, for a process with electron and positron beam polarizations,
the cross section can be expressed as \cite{pol},
\begin{eqnarray}
\sigma=\frac{1}{4}(1-P_{e^+})(1+P_{e^-})\sigma_{-1+1}+\frac{1}{4}(1+P_{e^+})(1-P_{e^-})\sigma_{+1-1}.
\end{eqnarray}
Here $\sigma_{ab}$ represents the calculated cross section with fixed helicities $a$ for positron and $b$ for the electron.
$P_{e^-}$ is the electron beam and $P_{e^+}$ is the positron relative polarisation.
It may be noted that the considered process includes only a weak interaction.
Therefore, only left-handed electrons (right-handed positrons) should be taken into account because of the structure of the $We^{-}\nu_{e}$ ($We^{+}\bar{\nu_{e}}$) vertex. Hence, the left-polarized electron (right-polarized positron) beam would enhance the cross section. This effect can be seen in Figs.\ref{fig1}-\ref{fig6} for various polarization schemes. The total cross section of the $e^{+}e^{-}\rightarrow \nu_{e} W^{-} W^{+} \bar{\nu}_{e}$ process as a function of $\frac{f_{S0}}{\Lambda^{4}}$ coupling for $1.4$ and $3$ TeV center-of-mass energies are shown in Fig.\ref{fig1} and Fig.\ref{fig2}, respectively.  In Fig.\ref{fig3} and Fig.\ref{fig4} depict the total cross section depending on the anomalous $\frac{f_{S1}}{\Lambda^{4}}$ coupling at $\sqrt s =$1.4 TeV and $\sqrt s =$3 TeV energies, respectively. The total cross sections are plotted in Fig.\ref{fig5} and  Fig.\ref{fig6} with respect to $\frac{f_{T0}}{\Lambda^{4}}$ coupling for the center-of-mass energies of 1.4 TeV and 3 TeV, respectively.
As can be seen from Fig.\ref{fig1} and Fig.\ref{fig6}, the polarization ($P_{e^-}=$-$80\%$; $P_{e^+}=$+$60\%$) enhances the cross sections the most compared to the other considered to polarization schemes. The lowest points of the curves in Fig.\ref{fig1} and Fig.\ref{fig6} correspond the value of the SM cross section. As seen from the figures, the increase of the centre-of-mass energy leads to remarkable enhancement of the deviations from the SM. For instance, in Fig.\ref{fig1} cross section increases nearly by a factor of $10$ from the SM for $\frac{f_{S0}}{\Lambda^{4}}=1\times10^{-9}$ GeV$^{-4}$ at $1.4$ TeV. However, cross section increases nearly by a factor of $100$ at $3$ TeV for the same value for $\frac{f_{S0}}{\Lambda^{4}}$. On the other hand, the obtained cross sections are very sensitive to the anomalous parameters. For instance in Fig.\ref{fig2} cross section increases two orders of magnitude as  $\frac{f_{S1}}{\Lambda^{4}}$ increases from $0$ to $1\times10^{-9}$ GeV$^{-4}$. As seen from these figures, the total cross sections  depend on the center of mass energy. When the center-of-mass energies are changed from $1.4$ TeV to $3$ TeV, the cross sections nearly increase by a factor of $100$. The cross sections are almost symmetric with respect to change in the sign of anomalous couplings. Therefore, main contribution comes from the quadratic anomalous couplings terms.

In order to examine the sensitivity to the anomalous couplings, we use one parameter $\chi^2$ criterion without systematic error. The $\chi^2$  function is defined as follows,
\begin{eqnarray}
\chi^2=\left(\frac{\sigma_{SM}-\sigma_{NP}}{\sigma_{SM}\delta_{stat}}\right)^2
\end{eqnarray}
where $\sigma_{NP}$ is the total cross section including SM and new physics, $\delta_{stat}=1/\sqrt{N}$ is the statistical error, $N$ is the number of background  events $N=L_{int}\sigma_{SM}$ where $L_{int}$ is the integrated CLIC luminosity. We have obtained 95\% C.L. limits on the anomalous coupling parameters using this analyze method at $\sqrt{s}=1.4$ TeV and $3$ TeV for different integrated luminosity values and final polarization configurations.  Polarization
improves the sensitivity bound of anomalous parameters as seen from the Tables \ref{tab3}-\ref{tab8}. As we expected, the best limits are obtained for the $P_{e^{-}}=-80\%$; $P_{e^{+}}=60\%$ polarization state.

The expected best sensitivities on $\frac{f_{S0}}{\Lambda^{4}}$ and $\frac{f_{S1}}{\Lambda^{4}}$
couplings in Tables. \ref{tab3}-\ref{tab6} are far beyond the sensitivities of the LHC. Also, we observe that  our limits for $\frac{f_{S0}}{\Lambda^{4}}$ in Table \ref{tab3} at $\sqrt{s}=1.4$ TeV are competitive with the results in Ref. \cite{lhc1} and one order of magnitude better than the ones reported for the LHC with $L=100$ fb$^{-1}$ by Ref. \cite{lhc3}. Additionally, our limits for $\sqrt{s}=3$ TeV are at the same order of magnitude with the LHC results for the $L=3000$ fb$^{-1}$ and $\sqrt{s}=100$ TeV. These results can be seen if Table \ref{tab2} is compared to Table \ref{tab4}. Similar interpretations can be made for our bounds on $\frac{f_{S1}}{\Lambda^{4}}$ from Table \ref{tab5} and Table \ref{tab6}. In addition, we can see from Table \ref{tab7} the limits on $\frac{f_{T0}}{\Lambda^{4}}$ for  $\sqrt{s}=1.4$ TeV are very close to the LHC bounds for $5\sigma$ with $300$ fb$^{-1}$ for $\sqrt{s}=14$ TeV. However, the bounds for $\sqrt{s}=100$ TeV future hadron collider with $L=3000$ fb$^{-1}$ are better than our limits on $\frac{f_{T0}}{\Lambda^{4}}$ as stated from Table \ref{tab8}. In the meantime, the relation between linear and non-linear parametrization are showed in [4] as the following formations,

\begin{eqnarray}
\label{ww}
   &\alpha_4=\frac{f_{S0}}{\Lambda^4}\frac{\upsilon^4}{8}&  \\ \nonumber
   &\alpha_4+2\alpha_5=\frac{f_{S1}}{\Lambda^4}\frac{\upsilon^4}{8}&.
\end{eqnarray}

In reference \cite{fab} the authors study the non-linear parametrization through the processes $pp\rightarrow l^\pm \nu_l l^\pm \nu_l jj$ and $pp\rightarrow l^\pm \nu_l l^\mp \nu_l jj$ at the LHC as mentioned above. This process can isolate the $WWWW$ vertex. The best 95\% C.L. bounds they have found $\alpha_4=[-0.0011; 0.0016]$ and $\alpha_5=[-0.0022; 0.0016]$ for the $\sqrt{s}=14$ TeV and $L=3$ ab$^{-1}$. By using Eq.\ref{ww}, these limits can be converted to linear parametrization. The results obtained in this case are $\frac{f_{S0}}{\Lambda^4}=[-2.40\times10^{-12}; 3,50\times10^{-12}]$ GeV$^{-4}$ and $\frac{f_{S1}}{\Lambda^4}=[-1.20\times10^{-11}; 1.75\times 10^{-11}]$ GeV$^{-4}$.
As seen from the Table IV and VI, these limits are of the same order as our best limits for the $\frac{f_{S0}}{\Lambda^4}$, but the best limits we obtain for $\frac{f_{S1}}{\Lambda^4}$ are one order of magnitude better.
Additionaly, when the $10\%$ systematic uncertainty factor is taken into account, the sensitivity of the our best obtained bounds decrease by about one order magnitude for $\frac{f_{S0}}{\Lambda^4}$ , $\frac{f_{S1}}{\Lambda^4}$ and $3$ times decrease for $\frac{f_{T0}}{\Lambda^4}$ coupling.

\section{Conclusions}

The CLIC is a high energy collider which has TeV scale energy and very high luminosity.
Particularly, operating with its high energy and luminosity is extremely important
in order to investigate the anomalous $\frac{f_{S0}}{\Lambda^{4}}$, $\frac{f_{S1}}{\Lambda^{4}}$ and $\frac{f_{T0}}{\Lambda^{4}}$ couplings through the process $e^{+}e^{-}\rightarrow \nu_{e} W^{-} W^{+} \bar{\nu}_{e}$.
Since the anomalous couplings depend on energy strongly, the cross sections that contain these couplings would have momentum dependence than those of the SM. We have found that the contribution of the anomalous quartic couplings to the total cross section increases with increasing center-of-mass energy. Because of the structure of $We^{-}\nu_{e}$ ($We^{+}\bar{\nu_{e}}$) vertex, it is found that certain polarizations of the beam increases the cross sections. In this respect, we find the better sensitivity for the $P_{e^-}=-80\%$; $P_{e^+}=60\%$ polarization state.

As a result, the CLIC with very clean experimental conditions and being free from strong interactions with respect to LHC,
high colliding energy and very high luminosity has a potential advantage over the LHC in studying the anomalous $\frac{f_{S0}}{\Lambda^{4}}$, $\frac{f_{S1}}{\Lambda^{4}}$ and $\frac{f_{T0}}{\Lambda^{4}}$ couplings.

\textbf{Conflict of Interest}

The author declare that there is no conflict of interest regarding the publication of this paper.

\pagebreak

\pagebreak

\begin{table}
\caption{Quartic vertices modified by each dimension-8 operator are marked with X \cite{baak}.}
\label{tab1}
\begin{ruledtabular}
\begin{tabular}{cccccccccc}
&$WWWW$ &$WWZZ$ &$ZZZZ$  &$WWAZ$ &$WWAA$ & $ZZZA$ & $ZZAA$ & $ZAAA$ & $AAAA$ \\ \hline
$L_{S0},L_{S1}$ & $X$& $X$& $X$& $$& $$& $$& $$& $$& $$\\
$L_{M0},L_{M1},L_{M6},L_{M7}$ & $X$& $X$& $X$& $X$& $X$& $X$& $X$& $$& $$\\
$L_{M2},L_{M3},L_{M4},L_{M5}$ & $$& $X$& $X$& $X$& $X$& $X$& $X$& $$& $$\\
$L_{T0},L_{T1},L_{T2}$ & $X$& $X$& $X$& $X$& $X$& $X$& $X$& $X$& $X$\\
$L_{T5},L_{T6},L_{T7}$ & $$& $X$& $X$& $X$& $X$& $X$& $X$& $X$& $X$\\
$L_{T8},L_{T9}$ & $$& $$& $X$& $$& $$& $X$& $X$& $X$& $X$\\
\end{tabular}
\end{ruledtabular}
\end{table}

\begin{table}
\caption{Current sensitivity bounds on anomalous parameters for $95\%$ C.L. with $20.3$ fb$^{-1}$  for $\sqrt{s}=8$ TeV \cite{344}, $99\%$ C.L. with $100$ fb$^{-1}$ for $\sqrt{s}=14$ TeV \cite{lhc1}, $95\%$ C.L. with $100$ fb$^{-1}$  for $\sqrt{s}=14$ TeV \cite{lhc3}, $5\sigma$ with $300$ fb$^{-1}$ for $\sqrt{s}=14$ TeV \cite{lhc2} and $95\%$ C.L. with $3000$ fb$^{-1}$  for $\sqrt{s}=100$ TeV\cite{lhc3}.
\label{tab2}}
\begin{ruledtabular}
\begin{tabular}{ccccc}
$L(fb^{-1})$ &$\sqrt{s}$ (TeV)&$\frac{f_{S0}}{\Lambda^{4}}$(GeV$^{-4}) $&$\frac{f_{S1}}{\Lambda^{4}}$ (GeV$^{-4})$  &$\frac{f_{T0}}{\Lambda^{4}}$ (GeV$^{-4})$\\
\hline
$20.3$ \cite{344} &$8$&$[-0.13;0.18]\times 10^{-8}$&$[-0.25;0.31]\times10^{-8}$&$-$\\
$100$ \cite{lhc1} &$14$&$[-2.2;2.4]\times 10^{-11}$&$[-2.5;2.5]\times10^{-11}$&$-$\\
$100$ \cite{lhc3} &$14$&$[-1.8;1.8]\times 10^{-10}$&$[-2.7;2.8]\times10^{-10}$&$[-5.8;5.9]\times 10^{-13}$\\
$300$ \cite{lhc2} &$14$&$-$&$-$&$[-1.2;1.2]\times 10^{-12}$\\
$3000$ \cite{lhc3} &$100$&$[-2.9;3.0]\times 10^{-12}$&$[-1.3;1.1]\times10^{-12}$&$[-3.7;3.0]\times 10^{-15}$\\
\end{tabular}
\end{ruledtabular}
\end{table}

\begin{table}
\caption{Sensitivity of $f_{S0}/\Lambda^4$ at 95\% C.L. for $\sqrt{s}$=1.4 TeV in units of GeV$^{-4}$.
\label{tab3}}
\begin{ruledtabular}
\begin{tabular}{cccc}
$L(fb^{-1})$ &Unpolarized &$P_{e^-}=-80\%$; $P_{e^+}=0$  &$P_{e^-}=-80\%$; $P_{e^+}=60$\\
\hline
$50$&$[-10.02;9.67]\times10^{-11}$&$[-8.63;8.39]\times10^{-11}$&$[-7.72;7.70]\times10^{-11}$\\
$250$&$[-6.76;6.41]\times10^{-11}$&$[-5.81;5.57]\times10^{-11}$&$[-5.22;4.90]\times10^{-11}$\\
$500$&$[-5.71;5.36]\times10^{-11}$&$[-4.91;4.67]\times10^{-11}$&$[-4.41;4.09]\times10^{-11}$\\
$750$&$[-5.18;4.83]\times10^{-11}$&$[-4.44;4.21]\times10^{-11}$&$[-4.00;3.69]\times10^{-11}$\\
$1000$&$[-4.83;4.48]\times10^{-11}$&$[-4.14;3.90]\times10^{-11}$&$[-3.74;3.42]\times10^{-11}$\\
$1250$&$[-4.58;4.23]\times10^{-11}$&$[-3.93;3.69]\times10^{-11}$&$[-3.54;3.23]\times10^{-11}$\\
$1500$&$[-4.38;4.04]\times10^{-11}$&$[-3.76;3.52]\times10^{-11}$&$[-3.39;3.08]\times10^{-11}$\\
\end{tabular}
\end{ruledtabular}
\end{table}

\begin{table}
\caption{Sensitivity of $f_{S0}/\Lambda^4$ at 95\% C.L. for $\sqrt{s}$=3 TeV in units of GeV$^{-4}$.
\label{tab4}}
\begin{ruledtabular}
\begin{tabular}{cccc}
$L(fb^{-1})$ &Unpolarized &$P_{e^-}=-0.80\%$; $P_{e^+}=0$  &$P_{e^-}=-0.80\%$; $P_{e^+}=0.60$\\
\hline
$50$&$[-12.99;12.27]\times10^{-12}$&$[-11.26;10.56]\times10^{-12}$&$[-9.89;9.51]\times10^{-12}$\\
$250$&$[-8.81;8.09]\times10^{-12}$&$[-7.65;6.95]\times10^{-12}$&$[-6.68;6.29]\times10^{-12}$\\
$400$&$[-7.88;7.16]\times10^{-12}$&$[-6.84;6.14]\times10^{-12}$&$[-5.96;5.58]\times10^{-12}$\\
$800$&$[-6.69;5.96]\times10^{-12}$&$[-5.81;5.11]\times10^{-12}$&$[-5.04;4.66]\times10^{-12}$\\
$1200$&$[-6.08;5.36]\times10^{-12}$&$[-5.29;4.59]\times10^{-12}$&$[-4.56;4.19]\times10^{-12}$\\
$1600$&$[-5.69;4.96]\times10^{-12}$&$[-4.95;4.25]\times10^{-12}$&$[-4.27;3.89]\times10^{-12}$\\
$2000$&$[-5.40;4.67]\times10^{-12}$&$[-4.70;4.00]\times10^{-12}$&$[-4.05;3.67]\times10^{-12}$\\
\end{tabular}
\end{ruledtabular}
\end{table}

\begin{table}
\caption{Sensitivity of $f_{S1}/\Lambda^4$ at 95\% C.L. for $\sqrt{s}$=1.4 TeV in units of GeV$^{-4}$.
\label{tab5}}
\begin{ruledtabular}
\begin{tabular}{cccc}
$L(fb^{-1})$ &Unpolarized &$P_{e^-}=-0.80\%$; $P_{e^+}=0$  &$P_{e^-}=-0.80\%$; $P_{e^+}=0.60$\\
\hline
$50$&$[-6.82;6.62]\times10^{-11}$&$[-5.92;5.69]\times10^{-11}$&$[-5.26;5.07]\times10^{-11}$\\
$250$&$[-4.60;4.39]\times10^{-11}$&$[-3.99;3.77]\times10^{-11}$&$[-3.55;3.36]\times10^{-11}$\\
$500$&$[-3.88;3.68]\times10^{-11}$&$[-3.38;3.15]\times10^{-11}$&$[-2.99;2.81]\times10^{-11}$\\
$750$&$[-3.52;3.31]\times10^{-11}$&$[-3.06;2.84]\times10^{-11}$&$[-2.72;2.53]\times10^{-11}$\\
$1000$&$[-3.28;3.08]\times10^{-11}$&$[-2.86;2.64]\times10^{-11}$&$[-2.54;2.35]\times10^{-11}$\\
$1250$&$[-3.11;2.90]\times10^{-11}$&$[-2.71;2.49]\times10^{-11}$&$[-2.40;2.22]\times10^{-11}$\\
$1500$&$[-2.98;2.77]\times10^{-11}$&$[-2.59;2.37]\times10^{-11}$&$[-2.30;2.11]\times10^{-11}$\\
\end{tabular}
\end{ruledtabular}
\end{table}

\begin{table}
\caption{Sensitivity of $f_{S1}/\Lambda^4$ at 95\% C.L. for $\sqrt{s}$=3 TeV in units of GeV$^{-4}$.
\label{tab6}}
\begin{ruledtabular}
\begin{tabular}{cccc}
$L(fb^{-1})$ &Unpolarized &$P_{e^-}=-0.80\%$; $P_{e^+}=0$  &$P_{e^-}=-0.80\%$; $P_{e^+}=0.60$\\
\hline
$50$&$[-8.85;7.99]\times10^{-12}$&$[-7.62;6.87]\times10^{-12}$&$[-6.92;5.97]\times10^{-12}$\\
$250$&$[-5.72;5.16]\times10^{-12}$&$[-5.23;4.47]\times10^{-12}$&$[-4.80;3.85]\times10^{-12}$\\
$400$&$[-5.08;4.52]\times10^{-12}$&$[-4.70;3.94]\times10^{-12}$&$[-4.33;3.38]\times10^{-12}$\\
$800$&$[-4.25;3.69]\times10^{-12}$&$[-4.02;3.26]\times10^{-12}$&$[-3.73;2.77]\times10^{-12}$\\
$1200$&$[-3.82;3.26]\times10^{-12}$&$[-3.67;2.91]\times10^{-12}$&$[-3.42;2.47]\times10^{-12}$\\
$1600$&$[-3.54;2.98]\times10^{-12}$&$[-3.44;2.69]\times10^{-12}$&$[-3.22;2.27]\times10^{-12}$\\
$2000$&$[-3.33;2.77]\times10^{-12}$&$[-3.28;2.52]\times10^{-12}$&$[-3.08;2.12]\times10^{-12}$\\
\end{tabular}
\end{ruledtabular}
\end{table}

\begin{table}
\caption{Sensitivity of $f_{T0}/\Lambda^4$ at 95\% C.L. for $\sqrt{s}$=1.4 TeV in units of GeV$^{-4}$.
\label{tab7}}
\begin{ruledtabular}
\begin{tabular}{cccc}
$L(fb^{-1})$ &Unpolarized &$P_{e^-}=-0.80\%$; $P_e^{+}=0$  &$P_{e^-}=-0.80\%$; $P_{e^+}=0.60$\\
\hline
$50$&$[-4.86;3.25]\times10^{-12}$&$[-4.30;2.75]\times10^{-12}$&$[-4.04;2.38]\times10^{-12}$\\
$250$&$[-3.58;1.97]\times10^{-12}$&$[-3.20;1.65]\times10^{-12}$&$[-3.07;1.40]\times10^{-12}$\\
$500$&$[-3.18;1.57]\times10^{-12}$&$[-2.86;1.31]\times10^{-12}$&$[-2.77;1.10]\times10^{-12}$\\
$750$&$[-2.98;1.37]\times10^{-12}$&$[-2.69;1.14]\times10^{-12}$&$[-2.61;0.95]\times10^{-12}$\\
$1000$&$[-2.85;1.24]\times10^{-12}$&$[-2.58;1.03]\times10^{-12}$&$[-2.52;0.85]\times10^{-12}$\\
$1250$&$[-2.76;1.15]\times10^{-12}$&$[-2.50;0.95]\times10^{-12}$&$[-2.45;0.78]\times10^{-12}$\\
$1500$&$[-2.68;1.08]\times10^{-12}$&$[-2.44;0.89]\times10^{-12}$&$[-2.40;0.73]\times10^{-12}$\\
\end{tabular}
\end{ruledtabular}
\end{table}

\begin{table}
\caption{Sensitivity of $f_{T0}/\Lambda^4$ at 95\% C.L. for $\sqrt{s}$=3 TeV in units of GeV$^{-4}$.
\label{tab8}}
\begin{ruledtabular}
\begin{tabular}{cccc}
$L(fb^{-1})$ &Unpolarized &$P_{e^-}=-0.80\%$; $P_e^{+}=0$  &$P_{e^-}=-0.80\%$; $P_{e^+}=0.60$\\
\hline
$50$&$[-4.44;3.09]\times10^{-13}$&$[-3.91;2.59]\times10^{-13}$&$[-3.58;2.24]\times10^{-13}$\\
$250$&$[-3.24;1.90]\times10^{-13}$&$[-2.89;1.57]\times10^{-13}$&$[-2.68;1.34]\times10^{-13}$\\
$400$&$[-2.97;1.63]\times10^{-13}$&$[-6.84;6.14]\times10^{-13}$&$[-2.48;1.14]\times10^{-13}$\\
$800$&$[-2.64;1.30]\times10^{-13}$&$[-2.38;1.07]\times10^{-13}$&$[-2.24;0.90]\times10^{-13}$\\
$1200$&$[-2.47;1.13]\times10^{-13}$&$[-2.24;0.92]\times10^{-13}$&$[-2.11;0.78]\times10^{-13}$\\
$1600$&$[-2.37;1.03]\times10^{-13}$&$[-2.15;0.83]\times10^{-13}$&$[-2.04;0.70]\times10^{-13}$\\
$2000$&$[-2.29;0.95]\times10^{-13}$&$[-2.09;0.77]\times10^{-13}$&$[-1.98;0.64]\times10^{-13}$\\
\end{tabular}
\end{ruledtabular}
\end{table}

\end{document}